\documentstyle[12pt,epsf]{article}

\def\be{\begin{equation}}
\def\ee{\end{equation}}
\def\bea{\begin{eqnarray}}
\def\eea{\end{eqnarray}}

\begin{document}

\begin{flushright}
hep-th/0002021
\end{flushright}

\pagestyle{plain}

\def\t{\tilde}
\def\cs{\frac{1}{4(2\pi\alpha')^2}}
\def\CV{{\cal{V}}}
\def\e{{\rm e}}
\def\haf{{\frac{1}{2}}}
\def\tr{{\rm Tr}}
\def\goes{\rightarrow}
\def\goal{\alpha'\rightarrow 0}
\def\gym{g_{{}_{YM}}}
\def\hphi{{\hat\phi}}

\begin{center}
{\Large {\bf D0-Branes As Light-Front Confined Quarks}}

\vspace{.2cm}

Amir H. Fatollahi 

\vspace{.2 cm}

{\it Institute for Advanced Studies in Basic Sciences (IASBS),}\\
{\it P.O.Box 45195-159, Zanjan, IRAN}

\vspace{.1cm}
{\it and}
\vspace{.1cm}

{\it Institute for Studies in Theoretical Physics and Mathematics (IPM),}\\
{\it P.O.Box 19395-5531, Tehran, IRAN}\\


{\sl fath@theory.ipm.ac.ir}
\end{center}

\begin{abstract}
We argue that different aspects of Light-Front QCD at confined
phase can be recovered by the Matrix Quantum Mechanics of D0-branes.
The concerning Matrix Quantum Mechanics is obtained from dimensional
reduction of pure Yang-Mills theory to 0+1 dimension. The aspects of
QCD dynamics which are studied in correspondence with D0-branes are:
1) phenomenological inter-quark potentials, 2) whiteness of hadrons
and 3) scattering amplitudes. In addition, some other issues such as the
large-$N$ behaviour, the gravity--gauge theory relation and also a
possible justification for involving ``non-commutative coordinates''
in a study of QCD bound-states are discussed.
\end{abstract}


\section{Introduction}
The idea of string theoretic description of gauge theories is an old one
\cite{Po3} \cite{Polya}. Despite of the years passed on this idea, it is still
activating different research works in theoretical physics
\cite{Polyakov}\cite{Pes}\cite{mvpol}\cite{adscft}. On the other hand, in the last
years our understanding about string theory is changed dramatically; a stream
which is usually called the ``second string revolution'' \cite{vafa}. The aim of
this stream is formulating of a unified string theory as a fundamental theory of
the known interactions.  One of the most applicable tools in the above program are
D$p$-branes \cite {Po1,Po2}. It is conjectured that D$p$-branes are perturbative
representation of nonperturbative (strongly coupled) string theories. 

It has been known for a long time that hadron-hadron scattering processes have two
different behaviours depending on the amount of momentum transfer
\cite{Close,roberts}. At large momentum transfer interactions appear as
interactions between the hadron constituents, partons or quarks, and some
qualitative similarities to electron-hadron scattering emerge.  At high energies
and small momentum transfers Regge trajectories are exchanged. Regge trajectories
provide a motivation for the first stringy picture of strong interaction. However,
the good fitting between the Regge trajectories and the mass of strong
bound-states is yet unexplained \cite{Po3,silas} . 

Deducing the apparently different observations above discussed from a unified
picture is the challenge of theoretical physics and it is tempting to search for
the application of the recent string theoretic progresses in this area. In this
way one may find D$p$-branes good tools whose dynamics may be taken as a proper
effective theory for the bound-states of quarks and QCD-strings (QCD electric
fluxes). To use the string theory tools for QCD-strings one should replace the
string theory parameters with those of QCD in a proper way.  The case here is in
the reverse direction of going from early days of string theory, as the theory of
strong interaction, to string theory, as the theory of gravity. 

To push the above idea, in two works \cite{9902,9905}, taking the dynamics of
D0-branes as a toy model, the potential and the scattering amplitude of two
D0-branes were calculated.  It is found that the potential between static
D0-branes is a linear potential \cite{lucha,rosner,collins,ansar}. Also the
potential between two fast decaying D0-branes, which in the extreme limit see each
other instantaneously, is calculated and the general results are found in
agreement with phenomenology \cite{lucha,rosner,ansar}. The scattering amplitude
of two D0-branes was calculated in \cite{9905} based on the results of \cite{Fat},
it is shown that the cross section shows the Regge pole-expansion. Regge behaviour
has been used some years ago to fit the hadron-hadron total cross section data
successfully \cite{DL,HERA} (see also \cite{222,344,312,445} for some more recent
application of this behaviour). 

Based on the results of \cite{9902}\cite{9905} and some further discussions, we
argue that different aspects of Light-Front formulation of QCD may be recovered by
the Matrix Quantum Mechanics of D0-branes.  In this paper we consider the Matrix
Quantum Mechanics resulting from dimensional reduction of $d+1$ dimensional pure
$U(N)$ YM theory to $0+1$ dimension. A detailed procedure of constructing this
matrix mechanics is presented in \cite{kaku}. In analogy with string theory ($d=9$
or 25), we call {\it D0-branes} the free-particles sector of the moduli space. We
hope that these kinds of studies shed light on possible new relation between
D-brane dynamics and gauge theories. Also we adjust our discussions to be in a
reasonable contact with the known phenomenological aspects, though the exact match
with experiments should not be expected at this level.

In Sec.2 we review the distinguished role of Light-Front coordinates for
explaining the scaling behaviour of hadrons structure functions; the same behaviour
which is taken as the consequence of point-like substructure in hadrons. In Sec.3
a short review of Matrix Quantum Mechanics of D0-branes is presented. In Sec.4 the
calculation of the inter D0-branes potential will be presented. The discussion on
the ``whiteness'' of D0-branes bound-states is given in Sec.5. In Sec.6 we deal
with the problem of scattering. Sec.7 is devoted to discussions. Three issues are
discussed in Sec.7: 1) large-$N$ limit, 2) quarks, gauge theory and gravity
solutions relation and 3) non-commutativity. The discussion on the
non-commutativity is on a possible justification for appearance
``non-commutative'' coordinates in the study of ``non-Abelian'' bound-states, such
as bound-states of quarks and gluons.

\section{QCD, Light-Cone And Constituent Quark Picture}
Before gauge theoretic description of strong interaction, QCD,
there was Constituent Quark Model (CQM) for hadrons.  According to CQM a
meson is just a quark-antiquark bound-state and a baryon is a three-quark
one. The bound-state problem has been extensively studied for years by
phenomenological inter-quark potentials to calculate various low-energy
quantities. The agreement between calculated and observed quantities has
been always too well to justify pursuing this approach to study hadron
properties \cite{lucha}.

Presently QCD is established to be the underlying theory for
strong bound-states and also it has been understood that
QCD-vacuum is a very complicated medium. In low energy the
coupling constant is large and so quantum fluctuations are highly
excited.  It means that basically ``sea'' of quarks and gluons
have considerable contribution to the properties of hadrons. Moreover,
the phenomena like confinement is believed to be  direct
consequences of the complex nature of the QCD-vacuum. So it seems
that hadron picture of QCD is not reconcilable with any few-body
picture of hadrons, like CQM (see \cite{wilson} for a good
discussion on this point).

Experimentally, substructure of hadrons is probed in sufficiently
large momentum transfer scatterings of a fundamental particle,
e.g.  an electron, in the so-called Deep Inelastic Scattering
(DIS)  experiments. The existence of a point-like substructure,
parton or quark, is taken as the reason for ``scaling'' behaviour
of hadron structure-functions, i.e. the absence of any ``scale''
is the consequence of point-like objects \cite{Close}. Along
the Bjorken's argument, and as we recall it in below, this scaling
behaviour has a simple interpretation in Light-Cone point of view
on the processes which are involved in DIS.  The story is
the same for Feynman's parton picture of DIS experiment and the
Light-Cone Frame's cousin, the Infinite Momentum Frame (IMF)
\cite{kogut}. By this simple interpretation of scaling in
Light-Cone Frame we hopefully  have a constituent
picture for hadrons reconcilable with QCD, and it is the reason
for developing the Light-Cone formulation of QCD during the past years
\cite{wilson,lc1,lc2}.

\begin{figure}[t]
\begin{center}
\leavevmode
\epsfxsize=60mm
\epsfysize=50mm
\epsfbox{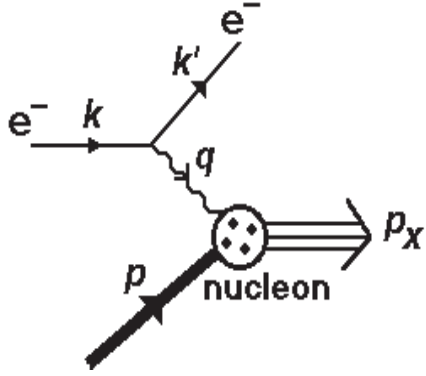}
\caption{{\it The lowest order process of DIS experiment.}}
\end{center}
\end{figure}

The unpolarized cross-section of DIS in the lowest order is given
by \footnote{This discussion is borrowed from \cite{roberts}
 and \cite{lcbook}.}
\bea
k'_0\frac{d\sigma}{d^3k'}=\frac{2M}{s-M^2}
\frac{\alpha^2}{Q^4} l_{\mu\nu}W^{\mu\nu}, \eea with \bea
W_{\mu\nu}(p,q)&=&\frac{1}{4M}\sum_\sigma \int \frac{d^4y}{2\pi}
\e^{iq\cdot y} \langle
p,\sigma|[J_\mu(y),J_\nu(0)]|p,\sigma\rangle,\\
l_{\mu\nu}&=&2(k_\mu k'_\nu+k_\nu k'_\mu -
\frac{1}{2}Q^2\eta_{\mu\nu}), \;\;q=k-k',\;q^2=-Q^2<0.
\eea
$M$ and $s$ are the mass of the nucleon and total energy respectively.
The momenta are specified in the Fig.1.  Also, we define the
useful parameters,
\bea
\nu=\frac{p\cdot q}{M},\;\;
x=\frac{Q^2}{2M\nu},\;\;y=\frac{2M\nu}{s-M^2}.
\eea
Note that parameters $x$ and $y$ are dimensionless. In the rest frame of
nucleon (target) we choose the z axis to be along
the virtual photon momentum then we have
\bea
p=(M,0,0,0),\;\;\;\;\;q=(\nu,0,0,-\sqrt{\nu^2+Q^2}).
\eea
In the so-called Bjorken limit, $Q^2\rightarrow\infty$,
$\nu\rightarrow\infty$ and $x$=fixed, we have
$q=(\nu,0,0,-\nu-Mx)$.  Now the statement of Bjorken scaling is as
following: Up to a kinematical coefficient, the hadronic tensor
$W_{\mu\nu}$ depends only on the parameter $x$ and not on
$Q^2$. To see this, it is convenient to use Light-Cone variables
$a^{\pm}=(a^0\pm a^3)/\sqrt{2}$ with scalar product as $a\cdot
b=a^+b^-+a^-b^+-a_T\cdot b_T$. Thus one writes
\bea\label{osc}
W_{\mu\nu}\sim \frac{1}{4M} \int dy^- \e^{iq^+y^-} \int
dy^+d^2y_T\e^{iq^-y^+}
 \langle p|J_\mu(y)J_\nu(0)|p\rangle.
\eea In the Bjorken limit we have \bea q^+\rightarrow
-Mx/\sqrt{2}={\rm fixed},\;\;\;\;q^-=(2\nu+Mx)/\sqrt{2}
\rightarrow \sqrt{2}\nu\rightarrow \infty.
\eea
In this limit the
integrand of (\ref{osc}) contains the rapidly oscillating factor
$\exp{(iq^-y^+)}$ which kills all contributions to the integral
except for those where the integrand is singular. Indeed the
singularity of integrand comes from the current product at
$y^+\sim 0$.  In addition due to causality the integrand vanishes for
$y^2=2y^+y^--y_T^2<0$.  So the dominant part of the integral comes
from $y^+=y_T=0$. It explains the Bjorken scaling, i.e., the $q^-$
no longer exists at $y^+=0$. Now it is clear
that the Light-Front coordinates play a distinguished role in the
understanding of the scaling behaviour in DIS experiments. The same result is
also correct for Feynman's parton description of DIS and IMF, the
experimental realization of Light-Cone Frame \cite{kogut}.

\section{Matrix Quantum Mechanics Of D0-Branes}
\setcounter{equation}{0}
According to string theory, D$p$-branes are $p$ dimensional objects
defined as (hyper)surfaces which can trap the ends of strings \cite{Po2} and
therefore it is reasonable to take their dynamics as a proper effective theory
for the bound-states of quarks and QCD-strings (QCD electric fluxes).

One of the most interesting aspects of D-brane dynamics appears in
their {\it coincident limit}. In the case of coinciding $N$
D$p$-branes their dynamics are captured by a $U(N)$ YM theory
dimensionally reduced to $p+1$ dimensions of D$p$-brane
world-volume \cite{W,Po2,Tay}. In the case of D0-branes $p=0$, the
above dynamics reduces to quantum mechanics of matrices, because
time is the only parameter in the world-line. A detailed procedure
of constructing this matrix mechanics is presented in \cite{kaku}.
The bosonic Lagrangian resulted from the pure YM is \cite{KPSDF}
\footnote{Here we take $d$ arbitrary.}
\bea\label{1.1}
&~&L=m_0 \tr\; \biggl(\haf  D_tX_i^2 + \cs[X_i,X_j]^2\biggl),\\
&~&i,j=1,...,d,\;\;\;\;\;\; D_t=\partial_t-i[a_0,\;],\nonumber
\eea
where $\frac{1}{2\pi\alpha'}$ and $m_0=(l_sg_s)^{-1}$ are the
string tension and the mass of D0-branes respectively
($l_s=\sqrt{\alpha'}$ and $g_s$ are the string length and
coupling, respectively). For $N$ D0-branes $X$'s are in adjoint
representation of $U(N)$ and have the usual expansion
$X_i=x_{i(a)} T_{(a)}$, $(a)=1,..., N^2$, \footnote{To avoid
confusion we put the group indices in ( ) always.}.

The action (\ref{1.1}) is invariant under the residual gauge symmetry of
unreduced YM theory. The transformations are:
\bea\label{gst1}
\vec{X}&\rightarrow&\vec{X'}=U\vec{X}U^\dagger,\nonumber\\
a_0&\rightarrow&a'_0=Ua_0U^\dagger+iU\partial_tU^\dagger,
\eea
where $U$ is an arbitrary time-dependent $N\times N$ unitary matrix. Under
these transformations one can check that:
\bea\label{gst2}
D_t\vec{X}&\rightarrow&D'_t\vec{X'}=U(D_t\vec{X})U^\dagger,\nonumber\\
D_tD_t\vec{X}&\rightarrow&D'_tD'_t\vec{X'}=U(D_tD_t\vec{X})U^\dagger.
\eea

First let us search for D0-branes in the above Lagrangian:\\
For each direction $i$ there are $N^2$ variables and not $N$ which one
expects for $N$ particles. However there is an ansatz for the equations of
motion which restricts the $U(N)$ basis to its $N$ dimensional Cartan
subalgebra. This ansatz causes vanishing the potential and one finds the
action of $N$ free particles, namely:
\bea\label{1.5}
S=\int dt \sum_{(a)=1}^N \haf m_0 \dot{\vec{x}}_{(a)}^2.
\eea
In this case the $U(N)$ symmetry is broken to $U(1)^N$ and the
interpretation of $N$ remaining variables as the classical (relative)
positions of $N$ particles is meaningful. The center of mass of D0-branes
is represented by the trace of the $X$ matrices.

In the case of unbroken gauge symmetry the gauge transformations mix the
entries of matrices and the interpretation of positions for D0-branes
remains obscure \cite{Ba}. Even in this case the center of mass is
meaningful and the ambiguity about positions only remains for the
relative positions of D0-branes.  In unbroken phase the $N^2-N$ non-Cartan
elements of matrices have a stringy interpretation; they govern the
dynamics of low lying oscillations of strings stretched between D0-branes.

The dependences of energy eigenvalues and the size of bound-states
are notable. By the scalings \cite{KPSDF}
\bea
t&\rightarrow& g_s^{-1/3} t,\nonumber\\
a_0&\rightarrow& g_s^{1/3} a_0,\nonumber\\
X&\rightarrow& g_s^{1/3} X,
\eea
one finds the relevant energy and size scales as
\bea\label{scales}
E &\sim& g_s^{1/3}/l_s,\nonumber\\ l_{d+2}&=& g_s^{1/3}l_s.
\eea
The length scale $l_{d+2}$ should be the fundamental length scale of the
covariant $d+2$ dimensional theory whose Light-Cone formulation is
argued to be described by action (\ref{1.1}) with longitudinal
momentum as $m_0$ \cite{Fat}. So it is natural to assume in our
case that $l_{d+2}$ (for $d=2$) is the inverse of the 3+1
dimensional QCD mass scale, denoted by $\Lambda_{QCD}$
\footnote{Due to Light-Front interpretation,
our  $\Lambda_{QCD}$ differs from \cite{kaku}. There
$l_s\sim\sqrt{\alpha'}$ is taken as $\Lambda_{QCD}^{-1}$.}. In the
weak coupling $g_s\to 0$ ($m_0\gg l_s^{-1}$) one finds $l_{d+2}\ll
l_s$ which allows to treat the bound-states of finite number of
D0-branes as point-like objects in the transverse directions of
the Light-Cone Frame \footnote{Because we admit the discrete
longitudinal momentum, $m_0$, for finite $N$,
we are dealing with Discrete-Light-Cone-Quantization (DLCQ)
\cite{9704080}. We do not emphasize on this point later.},
and consequently one finds $m_0\cdot E \sim \frac{1}{l_{d+2}^2}$,
which shows the invariance under Lorentz transformation of this
combination. As we will see in Sec.6 the masses of the
intermediate states in the scattering amplitude appear as
$l_{d+2}^{-1}$.

\section{Known Potentials}
\setcounter{equation}{0}
To calculate the effective potential between D0-branes one should find the
effective action around a classical configuration.  This work can be done
by integrating over the quantum fluctuations in a path integral. For the
diagonal classical configurations, classical
representations of D0-branes, the quantum fluctuations which must be
integrated over are the off-diagonal entries.  This work is equivalent to
integrating over the oscillations of the strings stretched between
D0-branes. Because here we deal with a gauge theory, and our interest
is calculation around the classical field configuration, to obtain the
effective action, it is convenient to use the background field method \cite{Abb}.

To calculate the effective action we write (\ref{1.1}) in $d+1$
space-time dimensions in the form (in the units $2\pi\alpha'=1$ and after
the Wick rotation $t\rightarrow it$ and $a_0 \rightarrow -ia_0$)
\bea\label{3.1}
L&=&m_0\tr\left(\frac{1}{4}[X_\mu,X_\nu]^2\right),\;\;\;
\mu ,\nu =0,1,...,d,\nonumber\\
X_0&=&i \partial _t +a_0,\;\;\;\;S=\int L dt,
\eea
where $\mu$ and $\nu$ are summed over by the Euclidean metric.  The one-loop
effective action of (\ref{3.1}) has been calculated several times (e.g.
see the Appendix of \cite{IKKT}) and the result
can be expressed as
\bea\label{3.5}
(\int dt)\;V(X_\mu^{cl}) =\frac{1}{2}\tr\log\bigg(
P_\lambda^2\delta_{\mu\nu}-2iF_{\mu\nu}\bigg)-
\tr\log\bigg(P_\lambda^2\bigg),
\eea
with
$$
P_\mu*\equiv[X_\mu^{cl},*],\;\;
F_{\mu\nu}\;*\equiv[f_{\mu\nu},*],\;\;
f_{\mu\nu}\equiv[X_\mu^{cl},X_\nu^{cl}],
$$
and
\bea\label{3.10}
P_\lambda^2=-\partial_t^2+\sum_{i=1}^d P_i^2,
\eea
with the backgrounds $a_0^{cl}=0$.
The second term in (4.2) is due to the ghosts associated with gauge symmetry.

\subsection{Static Potential}
Here we calculate the potential between two D0-branes at rest.
The classical solution which represents two D0-branes in distance $r$
can be introduced as
\bea\label{3.15}
X_1^{cl}=\haf \left( \matrix{ r & 0 \cr 0 & -r } \right),\;\;
X_0^{cl}=i\partial_t \left( \matrix{ 1 & 0 \cr 0 & 1 } \right),\nonumber\\
a_0^{cl}=X_i^{cl}=0,\;\;\;i=2,...,d.
\eea
So one finds
\bea\label{3.20}
P_1=\frac{r}{2}\otimes \Sigma_3,\;\;P_0=i\partial_t\otimes 1_4,\;\;
P_i=0,\;\;\;i=2,...,d,
\eea
where $\Sigma_3$ is the adjoint representation of the third Pauli matrix ,
$\Sigma_3*=[\sigma_3,*]$. The eigenvalues of $\Sigma_3$ are 0, 0, $\pm 2$.

The operator $P_\lambda^2$ is found to be
\bea\label{3.25}
P_\lambda^2=-\partial_t^2\otimes 1_4 + \frac{r^2}{4}\otimes
\Sigma_3^2,
\eea
which is a harmonic oscillator operator whose
frequency, reintroducing $\alpha'$, is $\omega\sim r/\alpha'$.
The one-loop effective action can be computed
\footnote{The one-loop effective action is a good approximation
for $\omega\gg m_0\dot{r}^2$. It gives $rg_s\gg l_s\dot{r}^2$
which for $g_s\rightarrow 0$ ($m_0\gg l_s^{-1}$) is satisfied for
large separations and low velocities.}.
\bea\label{a}
V(r) &=& (\frac{d-1}{2})\tr\log\bigg(P_\lambda^2\bigg) \nonumber\\
&=&- \;2(\frac{d-1}{2}) \int_0^\infty
\frac{ds}{s}\int_{-\infty}^{\infty} dk_0\;
\e^{-s(k_0^2+r^2)}\nonumber\\ &~&+\; {\rm traces\; independent\;
of\; } r,
\eea
where 2 is for the degeneracy in eigenvalue 4 of
$\Sigma_3^2$, and $k_0$ is for the eigenvalues of the operator
$i\partial_t$.  In writing the second line we have used
$$
\ln \bigg(\frac{u}{v}\bigg)=\int_0^\infty \frac{ds}{s} \;
(\e^{-sv}-\e^{-su}).
$$
The integrations can be performed and one finds
\bea\label{3.30}
V(r)&=&-\; 2(\frac{d-1}{2}) \int_0^\infty
\frac{ds}{s} (\frac{\pi}{s})^{\haf} \e^{-sr^2} \nonumber\\ &=&\;4
\pi (\frac{d-1}{2})\; |r|\; - \;
 \infty\; ({\rm independent \;of\;} r).
\eea
The linear potential is the same of phenomenology interests, see e.g.
\cite{lucha,collins,ansar}. Also it is the same which is consistent with
spin-mass Regge trajectories \cite{lucha,rosner,collins,ansar}.  By
restoring the $\alpha'$ the potential will be found to be
\bea\label{3.35}
V(r)=4 \pi (\frac{d-1}{2}) \frac{|r|}{2\pi\alpha'}
\eea
which has the dimension $length^{-1}$.  By comparison with Regge model one
can have an estimation for $\alpha'$ \cite{rosner,ansar}.
The above potential can be used to describe an effective theory
for the relative dynamics of D0-branes as
\bea\label{3.36}
S=\int dt \bigg(\frac{1}{2} \frac{m_0}{2} \dot{\vec{r}}^2-
4 \pi (\frac{d-1}{2}) \frac{|\vec{r}|}{2\pi\alpha'}\bigg),
\eea
which in the range of validity of one-loop approximation, mentioned in previous
footnote, it is expected to be applicable. Also by this action
one obtains the energy scale as $E\sim \alpha'^{-2/3}m_0^{-1/3}\sim
g_s^{1/3}/l_s$, as pointed in (\ref{scales}). The above action describes the
dynamics in Light-Cone Frame with the longitudinal momentum $m_0$, and
recalling (\ref{scales}) we have
$p^+p^-\sim m_0E\sim g_s^{-2/3}l_s^{-2}\sim l_{d+2}^{-2}$.

It is not hard to see that the two D0-brane interaction potential
is also true for every pair inside a bound-states of D0-branes. So
the effective action for $N$ D0-branes is found to be
\bea
S=\int dt \bigg(\frac{1}{2} m_0\sum_{(a)=1}^{N} \dot{\vec{r}}_{(a)}^2-
4 \pi (\frac{d-1}{2}) \sum_{(a)>(b)=1}^{N}
\frac{|\vec{r}_{(a)}-\vec{r}_{(b)}|}{2\pi\alpha'}\bigg).
\eea
In a recent work \cite{krishna}, by taking the linear potential between
quarks of a baryonic state in transverse directions of Light-Cone Frame, the
structure functions are obtained with a good agreement with observed ones.

It is useful to relate the parameter $1/\alpha'$ in the potential
with gauge theory parameters. To do this we need a string
theoretic description of gauge theory, but in the Light-Cone
Frame. The nearest formulation we know for this is
Light-Cone--lattice gauge theory (LClgt) \cite{lclgt}. In LClgt
one assumes time direction and one of the spatial directions, say
$z$, in continuum limit. The Light-Cone variables are defined as
usual $x^{\pm}\sim t\pm z$. Other spatial directions naturally
play the role of transverse directions of Light-Cone Frame, which
are assumed to be on a lattice in LClgt. Due to existence of a
continuous time $x^+$, there exists a Hamiltonian formulation
\cite{kosu} of the lattice gauge theory \cite{willat}. The relation
between the linear confinement potential and gauge-lattice
parameters is given by \cite{kosu,lclgt}:
\bea
V(r)\sim \frac{\gym^2}{a^2} |r|,
\eea
with $a$ as the lattice spacing
parameter in the transverse directions. Comparing this with (4.9)
leads to \bea \frac{1}{\alpha'}\sim \frac{\gym^2}{a^2}. \eea
\subsection{Fast Decaying D0-Branes}
\footnote{This subsection was modified based on the crucial comment by refree
of EPJ.C.}
For two {\it fast decaying} D0-branes one can again calculate the above
potential. This work can be done by inserting for example a Gaussian function
for $k_0$ into the (\ref{a}).  This work is equivalent to restricting the
eigenvalues of the operator $i\partial_t$. Having this in mind that
eigenvalues of operators ($X,\;i\partial_t$, ...) represent the
information corresponding to classical values of D0-branes space-time
positions
\footnote{The eigenvalues of $i\partial_t$ here are different from their
quantum mechanical analogue which due to the Schrodinger's equation, are
energy.}, we find
\bea\label{3.40}
V(r)&=&-2(\frac{d-1}{2}) \int_0^\infty \frac{ds}{s}
\int_{-\infty}^{\infty} dk_0\;
\bigg(\frac{1}{\Delta} \e^{\frac{-k_0^2}{\Delta^2}}\bigg)
\e^{-s(k_0^2+r^2)}\\
&=&
-2 \sqrt{\pi}(\frac{d-1}{2}) \int_0^\infty \frac{ds}{s} 
\frac{e^{-sr^2}}{\sqrt{s\Delta^2+1}},
\eea
in which we assumed that the D0-branes live around time zero. The
last expression is infinite, but one can show that the infinite part is
$r$-independent. One takes:
\bea
\frac{\partial V(r)}{\partial (r^2)}= 
2 \sqrt{\pi}
(\frac{d-1}{2}) \int_0^\infty \frac{ds 
e^{-sr^2}}{\sqrt{s\Delta^2+1}},
\eea
which is finite and so infinity of $V(r)$ is $r$-independent. The last 
integral can not be calculated exactly, though numerical comparison with
phenomenology is possible. The limit $\Delta\rightarrow 0$ can be calculated
exactly by recalling the relation:
$$
\lim_{\Delta\rightarrow0}
\bigg(\frac{1}{\Delta}e^{-\frac{k_0^2}{\Delta^2}}\bigg)=\sqrt{\pi}
\delta(k_0).
$$
Inserting $\delta$-function in (\ref{3.40}) one finds:
\bea\label{3.50}
V(r)\sim -2(\frac{d-1}{2}) \int_0^\infty
\frac{ds}{s}\; \e^{-s(r^2)}\;\sim\;\ln r,
\eea
which the last result is after extracting the $r$-independent infinity. This
result is already consistent with phenomenology of heavy quarks
\cite{ansar,rosner}, which we know their weak decay rates grow with $(mass)^5$.
In the extreme limit $\Delta \rightarrow 0$, in which the two D0-branes see
each other ``instantaneously'', one can take them as two D(-1)-branes
(D-instantons). The dynamics of D(-1)-branes are described by the action
(\ref{3.1}) but instead of the taking $X_0$ as $i\partial_t$ one takes $X_0$
as a matrix which its eigenvalues represent the ``instants'' which D(-1)-branes
occur. So the above logarithmic result also could be obtained in D(-1)-branes
calculation by taking a classical solution as
\bea\label{3.45}
X_1^{cl}=\haf \left( \matrix{ r & 0 \cr 0 & -r } \right),\;
X_0^{cl}=\left( \matrix{ t_0 & 0 \cr 0 & t_0 } \right),
\;\;a_0^{cl}=X_i^{cl}=0,\;\;i=2,...,d     ,
\eea
which represents two D(-1)-branes appeared at time $t_0$, in distance $r$.

A comment is in order: from phenomenological point of view, it is known that in
some cases potentials like $r^\xi$, with $\xi\simeq 0.1$ also have produced
good results \cite{ansar,rosner}.
This maybe can be included to our intermediate result (\ref{3.40})
or logarithmic result by recalling the
numerical relation $\ln r \simeq r^{\eta\simeq 0}$, which is valid for a range
of $r$
\footnote{One can justify by the relation:
$$
\ln r = \lim_{\eta\rightarrow 0} \int dr\;
r^{-1+\eta} = \lim_{\eta\rightarrow 0} r^\eta/\eta
$$}.

\section{White States}
\setcounter{equation}{0} To determine the color of an object its
dynamics should be studied in presence of external fields. For a
``white'' extended object, the center of mass (c.m.)  moves as a
free particle in a uniform electric field. Now we want to specify
the color of the D0-branes bound-states.  As we will see, although
our formulation for dynamics of D0-branes in external YM fields
seems incomplete, but there is a reasonable statement about
``whiteness'' of D0-branes bound-states.
\subsection{D0-Branes In YM Background}
In classical Electrodynamics besides electromagnetic fields produced
by different distributions of charges and currents, we also study the
dynamics of a charged particle in regions of space where electromagnetic
fields exist.  There is a simple question:
What are the problems arising when one studies Chromodynamics in this level?

The main problem arises when one introduces sources and
matches Chromodynamics with dynamics of colored objects (for
example a colored particle).  In case of Electrodynamics there is
a simple relation. For example the equation of motion of a charge
particle with mass $m_0$ and charge $q$ is
\bea\label{lf}
m_0\ddot{\vec{x}}=q(\vec{E}+\vec{v}\times \vec{B}).
\eea
The concept of gauge invariance at this level is understood as the
invariance of equations of motion under gauge transformations,
i.e. field strengths are invariant under gauge transformations.
Now, in the case of Chromodynamics right-hand-side is a matrix and
transforms as an object in adjoint representation under gauge
group transformations, as
\bea
\vec{E}\rightarrow \vec{E}'=U\vec{E} U^\dagger,\;\;\;\;
\vec{B}\rightarrow \vec{B}'=U\vec{B} U^\dagger.
\eea
So the problem arises.  As it is
well-known for string theorists, now we have a good candidate for
non-commutative coordinates which are the coordinates of
coincident D0-branes. First one may rewrite (5.1) for ``matrix''
coordinates as
\bea
m_0{\ddot{\vec{X}}}=q(\vec{E}+{\dot{\vec{X}}}\times \vec{B}),
\eea
but it is not enough to have correct behaviour for the first side
under gauge transformations. Here the world-line gauge symmetry
(\ref{gst1})  of D0-brane dynamics helps us, to write the
generalized Lorentz equation as \footnote{Here we drop the
commutator potential in the action of D0-branes, without any lose
of generality.}, \footnote{One may be easier with the symmetrized
version of the magnetic part as
$\frac{1}{2}(D_t\vec{X}\times\vec{B}-\vec{B} \times D_t\vec{X})$.}
\bea\label{gle}
m_0D_tD_t\vec{X}=q(\vec{E}+D_t\vec{X}\times
\vec{B}).
\eea
By recalling the relation (\ref{gst2}) one observes
that both sides have the same behaviour under gauge
transformations. However, it seems that the picture is not complete
yet. First, it is not clear what is the Lagrangian formulation
of this problem. Secondly, the precise meaning of position
dependences of field strengths should be
clarified (there is the same question for $U$, the parameter of
gauge transformation).

Now, the crucial observation is the decoupling
of c.m. dynamics from non-Abelian parts. It is
because of trace nature of $U(1)$ and $SU(N)$ parts.  As we mentioned
earlier the c.m. degree of freedom is described by the
$U(1)$ part of $U(N)$ \cite{W}. So the position and the momentum of c.m.
can be obtained by a simple trace \cite{Ba}
\bea\label{2.55}
\vec{x}_{c.m.}\equiv\frac{1}{N}\; \tr\vec{X},  \;\;\;\;\;
\vec{p}_{c.m.}\equiv \; \tr\vec{P}.
\eea
To investigate the kind and amount of the charge of an object its
dynamics should be
studied in absence of magnetic field ($\vec{B}=0$) and (for extended
objects) in uniform electric field ($\vec{E}(x)=\vec{E}_0$). So the c.m. equation
of motion is
\bea
m_0\ddot{\vec{x}}_{c.m.}=q\vec{E}_{(1)0},
\eea
\begin{figure}[t]
\begin{center}
\leavevmode
\epsfxsize=100mm
\epsfysize=50mm
\epsfbox{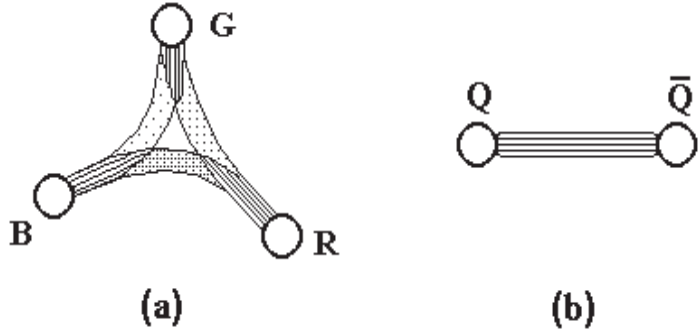}
\caption{{\it The net electric flux extracted from each
quark is equivalent in a baryon (a) and a meson (b).
The D0-brane--quark correspondence suggests the string-like
shape for fluxes inside a baryon (a).}}
\end{center}
\end{figure}
\noindent which the subscript {\small (1)}  denotes that the corresponding
electric field comes from the $U(1)$ part of $U(N)$. It is understood
that the dynamics of c.m. will not be affected by the non-Abelian
part of gauge group.  It means that the c.m. is white with respect
to $SU(N)$. This behaviour of D0-branes bound-states is the same as that
of hadrons. It means that each D0-brane feels the net effect of
other D0-branes as the white-complement of its color. In other
words, the field fluxes extracted from one D0-brane to the other ones
are the same as the flux between a color and an anti-color,
Fig.2.  As we have shown in Sec.4, there is a linear potential between
each two static D0-branes, which is consistent with this
flux-string picture. Also, the number of D0-branes in the
bound-state, $N$, equals to that of baryons. As we mentioned
before, recently \cite{krishna} the linear potential between the
constituents of baryons, in the transverse directions of
Light-Cone Frame, has been used successfully to obtain the
structure functions.

As the final note in this part, we remind that the dynamics
presented by (\ref{lf}) can be taken as for a massless particle in
transverse directions in Light-Cone Frame with longitudinal
momentum, $p^+\equiv m_0$. The fields $\vec{E}$ and $\vec{B}$
are electromagnetic fields in transverse directions. We present
the derivation of this in the Appendix.

\section{Scattering Amplitude}
\setcounter{equation}{0}
As a consequence of asymptotic freedom, in a suddenly collision process
quarks or partons are assumed to be free. So the probe, an electron or
another quark, only interacts with the hadron constituents instead of the hadron
as a whole \cite{Close,roberts}. It is the same mechanism which results {\it
scaling} behaviour in hadron structure functions.

With the above in mind it is reasonable to calculate the scattering
amplitude between two individual D0-branes, to find an impression
about the behaviour of the scattering amplitude of two hadrons which D0-branes are
assumed as their quarks. Also it is natural to assume that this result is
for high energy-elastic regime of hadron
collisions.

Here we use the result of \cite{Fat}. In \cite{Fat} it is shown that the
quantum travelling of D0-branes can be understood by the field theory
Feynman graphs and corresponding amplitudes in the Light-Cone Frame. In
the following we review the approach to calculate the amplitude.

We concentrate on the limit $\goal$. In this limit to have a
finite energy one has
\bea\label{4.1}
[X_i,X_j]=0,\;\;\;\forall\;i,j,
\eea
and consequently
the potential term in the action vanishes.  So, D0-branes do not interact
and the ``classical action'' reduces to the action of $N$ free
particles. We take this classical action also in the quantum case too,
it is equivalent to the assumption that two quarks in two spatially
well separated hadrons do not interact with each other. Since hadrons are
white one can trust this assumption. However, the
above observation fails when D0-branes arrive each other. When two
D0-branes come very near each other two eigenvalues of $X_i$
matrices will be equal and the corresponding off-diagonal elements
can get non-zero values. This is the same story of gauge symmetry
enhancement. The fluctuations of these off-diagonal elements are
responsible for the interaction between D0-branes in bound-states.

In the coincident limit the dynamics is complicated.
The relative matrix position may be taken as:
\bea\label{4.10}
\vec{X}= \left( \matrix{\vec{r}/2 & \vec{Y} \cr
\vec{Y}^* & -\vec{r}/2 }\right),
\eea
where $Y^*$ is the complex conjugate of $Y$. By inserting this matrix into
the Lagrangian one obtains:
\bea\label{4.15}
S&=&\int dt\haf \bigg( (2m_0) \dot{\vec{X}}_{c.m.}^2 +
m_0 \dot{\vec{Y}}\cdot\dot{\vec{Y}}^*
- \frac{m_0}{4} \cs (1-\cos^2\theta)
\vec{r}^{{}^2} \vec{Y}\cdot\vec{Y}^*\nonumber\\
&+&\frac{m_0}{2}\dot{\vec{r}}^2+O(Y^3)\bigg),
\eea
with $X_{c.m.}$ for the center of mass and $\theta$ is the angle between
$\vec{r}$ and the complex vector $\vec{Y}$.  As it is apparent in the
$\goal$ limit which is the case of our interest, the $r$ element do not
take large
values and have a small range of variation. In high-tension approximation
of strings ($\alpha'\rightarrow 0$), one can take the separation of
D0-branes
a constant of order $r\sim g_s^{1/3}l_s$. As is noted in Sec.3, this
length is the typical size of the D0-brane bound-states. So,
\bea\label{4.20}
S=\int dt \bigg(\haf (2m_0) \dot{\vec{X}}_{c.m.}^2 +
\haf m_0 \dot{Y}_\perp\cdot\dot{Y}_\perp^*
- \haf m_0 \frac{k^2r^2}{\alpha'^2}
Y_\perp\cdot Y_\perp^* +\haf\frac{m_0}{2}\dot{\vec{r}}+\cdot\cdot\cdot
\bigg),
\eea
where in the above $k$ is a numerical factor depending on $\alpha'$ and $g_s$
, and $Y_\perp$ is the part of the $\vec{Y}$ perpendicular to the
relative distance $\vec{r}$. The parallel part of $\vec{Y}$ behaves as a
free part. In $d+1$ dimensions of space-time the dimension of $Y_\perp$
is $d-1$ which shows that we are encountered with $2\times(d-1)$ harmonic
oscillators because, $Y$ is a complex variable. This is the same number
of harmonic oscillators which appears in one-loop calculations (Sec.4).
These harmonic oscillators correspond to vibrations of (oriented)
open strings stretched between D0-branes. In the
following we ignore the radial momentum and even
the angular momentum by dropping the term $m_0\dot{\vec{r}}^2$
and set $r=r_0$ for simplicity
\footnote{Setting $r=r_0$ may be justified by a mean value
problem in integrations over constant backgrounds in the path integral as:
$$
\int r^{d-1}dr \int DY\:DY^*\e^{-S[r,Y,Y^*]}\sim
\int DY\:DY^*\e^{-S[r_0,Y,Y^*]}.
$$}.

For two D0-branes we take the probability amplitude presented by
path integral as
\bea\label{4.25}
\langle x_3,x_4;t_f| x_1, x_2;t_i  \rangle=\int \e^{-S}.
\eea
Based on the previous discussion,  in the $\goal$ limit for (Fig.3) graph we
decompose the path-integral as the following,
\footnote{Here similar to what we have in field theory we
have dropped the dis-connected graphs.},
\bea\label{4.30}
&~&\langle x_3,x_4;t_f| x_1, x_2;t_i  \rangle=
\left[\int \e^{-S}\right]_{\goal}=
  \int_{t_i}^{t_f} dT_1  dT_2
  \int_{-\infty}^{\infty} dX_1 dX_2  \nonumber\\
&~&\times
\bigg(K_{m_0}(X_1,T_1;x_1,t_i) K_{m_0}(X_1,T_1;x_2,t_i) \bigg)    \nonumber\\
&~&\times
\bigg(K_{2m_0}(X_2,T_2;X_1,T_1)
K_{oscillator}(Y_\perp=0,T_2;Y_\perp=0,T_1)\bigg)
\nonumber\\
&~&\times
 \bigg(K_{m_0}(x_3,t_f;X_2,T_2) K_{m_0}(x_4,t_f;X_2,T_2)\bigg),
\eea
which $K_m(y_2,t_2;y_1,t_1)$ is the non-relativistic propagator of a free
particle with mass $m$ between $(y_1,t_1)$ and $(y_2,t_2)$ and
$K_{oscilator}(Y_\perp=0,T_2;Y_\perp=0,T_1)$ is the harmonic oscillator
propagator. $\int dT_1 dT_2 dX_1 dX_2$ is for a summation over different
``Joining-Splitting'' times and points. We use in $d$ dimensions the
representations
$$
K_{m}(y_2,t_2;y_1,t_1)=\theta(t_2-t_1)\frac{1}{(2\pi)^d} \int d^d p
\;\e^{ip\cdot(y_2-y_1)-\frac{ip^2(t_2-t_1)}{2m}},
$$
$$
K_{oscilator}(Y_\perp=0,T_2;Y_\perp=0,T_1)=
\theta(T_2-T_1)\bigg(\frac{m_0\omega}{2\pi i
\sin[\omega(T_2-T_1)]}\bigg)^{d-1},
$$
where $\theta(t_2-t_1)$ is the step function and
$\omega$ is the harmonic oscillator frequency,
$\omega\sim kr_0/\alpha'\sim kg_s^{1/3}/l_s$.
Because of complex nature of $Y_\perp$ the
power for the harmonic propagator is  $2\times\frac{d-1}{2}$.

\begin{figure}[t]
\begin{center}
\leavevmode
\epsfxsize=80mm
\epsfysize=80mm
\epsfbox{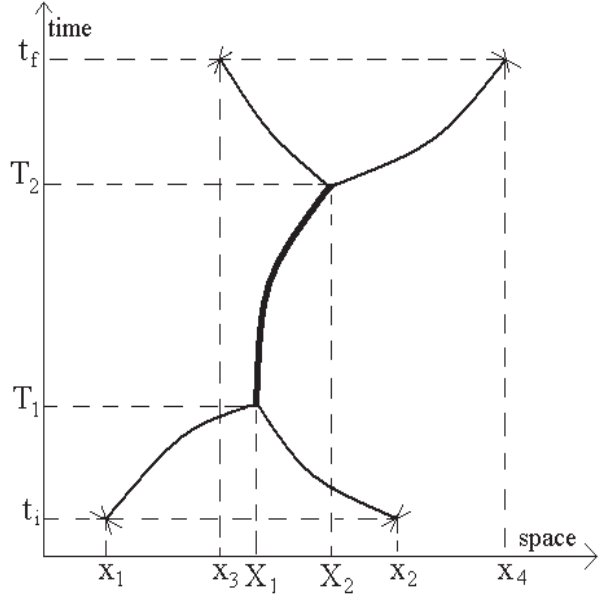}
\caption{{\it A typical tree path of D0-branes.}}
\end{center}
\end{figure}

All the above results can be translated into the momentum space
($E_k=\frac{p_k^2}{2m_0}$ with $k=1,2,3,4$):
\bea\label{4.32}
\langle p_3,p_4;t_f |p_1, p_2;t_i  \rangle&\sim&
\e^{i(E_3+E_4)t_f-i(E_1+E_2)t_i}
\int \prod_{a=1}^4 dx_a \e^{i(p_1x_1+p_2x_2-p_3x_3-p_4x_4)}
\nonumber\\&~&\times\langle x_3,x_4;t_f | x_1, x_2;t_i \rangle.
\eea
This representation is useful to calculate the cross section.
The integrals can be performed and we find
\bea
&~&\langle p_3,p_4;t_f |p_1, p_2;t_i\rangle            \sim
\delta^{(d)}(p_1+p_2-p_3-p_4)
\int_{t_i}^{t_f} dT_1 dT_2 \theta(T_2-T_1)
\nonumber\\
&~&\times\exp(\frac{-i(p_1^2+p_2^2)T_1}{2m_0})
\exp(\frac{-iq^2(T_2-T_1)}{4m_0})
\exp(\frac{i(p_3^2+p_4^2)T_2}{2m_0})
\nonumber\\
&~&
K_{oscillator}(Y_\perp=0,T_2;Y_\perp=0,T_1)
\eea
where $\vec{q}=\vec{p}_1+\vec{p}_2=\vec{p}_3+\vec{p}_4$.

To have a real scattering process let us assume $$ t_i\rightarrow
-\infty,\;\;\;t_f\rightarrow \infty. $$ We put $T\equiv T_2-T_1$
which has the range $0\leq T \leq \infty $. The integrals yield
\bea
\langle p_3,p_4;\infty |p_1, p_2;-\infty\rangle &\sim&
\delta^{(d)}(p_1+p_2-p_3-p_4)
\delta(\frac{p_1^2}{2m_0}+\frac{p_2^2}{2m_0}-
\frac{p_3^2}{2m_0}-\frac{p_4^2}{2m_0}) \nonumber\\ &~&
\int_0^\infty dT\; \e^{\frac{-iT}{4m_0}(q^2-2(p_1^2+p_2^2))}
\bigg(\frac{m_0\omega}{\sin (\omega T)}\bigg)^{d-1}.
\eea
Recalling the energy-momentum relation in the Light-Cone gauge
\cite{Fat},
$$
2(p_1^2+p_2^2)-\vec{q}^{\;2}=
2(2m_0)(\frac{p_1^2+p_2^2}{2m_0})- \vec{q}^{\;2}=
2q^+q^--\vec{q}^{\;2}=q_\mu q^\mu\equiv q_\mu^2,
$$
we find
\bea
\langle p_3,p_4,E_3,E_4;\infty
|p_1,p_2,E_1,E_2;-\infty\rangle &\sim&
\delta^{(d)}(p_1+p_2-p_3-p_4)
\delta(E_1+E_2-E_3-E_4)\nonumber\\&~& \int_0^\infty dT
\e^{\frac{-q_\mu^2}{4m_0}T} \bigg(\frac{m_0\omega}{\sin (\omega
T)}\bigg)^{d-1}.
\eea
We perform a cut-off for $T$ in small values
as $0 < \epsilon\leq T \leq \infty $, with $\epsilon$ be small
\footnote{This cut-off is for extracting the contribution of
graphs with four-legs vertex, as $\lambda \phi^4$. From  time-energy
uncertainty relation, we learn that these graphs are
generated by super-heavy intermediate states.}. By changing the
integral variables as $\e^{-2\omega T}=\eta$, we have
\bea\label{amp}
\langle p_3^\mu,p_4^\mu;\infty
|p_1^\mu,p_2^\mu;-\infty\rangle &\sim&
\delta^{(d)}(p_1+p_2-p_3-p_4)
\delta(p_1^-+p_2^--p_3^--p_4^-)\nonumber\\&~&
\frac{(m_0\omega)^{d-1}}{2\omega} \int_0^x d\eta\;
\eta^{\frac{-q_\mu^2}{8m_0\omega}+\frac{d-3}{2}}
(1-\eta)^{-d+1},\nonumber\\ &\sim& \delta^{(d)}(p_1+p_2-p_3-p_4)
\delta(p_1^-+p_2^--p_3^--p_4^-) \nonumber\\&~&
\frac{(m_0\omega)^{d-1}}{2\omega}
B_x(\frac{-q_\mu^2}{8m_0\omega}+\frac{d-1}{2},-d+2)
\eea
where $1\sim x=\e^{-2\omega\epsilon}$ and $B_x$ is the Incomplete Beta
function. The longitudinal momentum conservation trivially is
satisfied. Furthermore, because of the conservation of this momentum we
do not expect so-called $t$-channel processes here.

\subsection{Polology}
Equivalently one may use the other representation of $K_{oscillator}$
as
\bea
K_{oscillator}(Y_\perp=0,T_2;Y_\perp=0,T_1)=
\sum_{n} \langle0|n\rangle\langle n|0\rangle \e^{-iE_n(T_2-T_1)},
\eea
with $E_n$'s as the known $H_{oscillator}$ eigenvalues. In this
representation one finds the pole expansion \cite{Fat}:
\bea
\langle p_3^\mu,p_4^\mu;\infty |p_1^\mu, p_2^\mu;-\infty  \rangle&\sim&
\delta^{(d)}(p_1+p_2-p_3-p_4)
\delta(p_1^-+p_2^--p_3^--p_4^-)
\nonumber\\
&~&\times \lim_{\epsilon \rightarrow 0^+}
\sum_n C_n\; \frac{i4m_0}{ q_\mu q^\mu - M_n^2+i\epsilon}.
\eea

This pole expansion also can be derived by extracting the poles
of the amplitude (\ref{amp}) with the condition
\bea
\frac{-q_\mu^2}{8m_0\omega}+\frac{d-1}{2}=-n, \;\;
{\rm with}\; n\; {\rm as\; a\; positive\; integer}.
\eea
Hence for the mass of the intermediate
bound-states we obtain
\bea\label{Regge}
M_n^2=\frac{8k(n+\frac{d-1}{2})}{(g_s^{1/3}l_s)^2}.
\eea
We recall that the combination $g_s^{1/3}l_s$ is $l_{d+2}$,
the fundamental length of $d+2$ dimensional theory (Sec.3 and \cite{Fat}).
The Regge pole-expansion of (\ref{amp})-(\ref{Regge})
is the phenomenological promising
feature of this amplitude \cite{DL,HERA,222,344,312,445}.

\section{Discussion}
\setcounter{equation}{0}
In this section we discuss some relevant issues: 1) large-$N$ limit, 2)
quark, gauge theory and gravity solutions relations and also
3) non-commutativity.

\subsection{Large-$N$}
Baryons show special properties in large-$N$ limit of gauge theories
\cite{witN}
\begin{itemize}
\item Their mass grows linearly by $N$.
\item Their size do not depend on $N$. So their density goes to
infinity at large-$N$.
\item Baryon-baryon force grows proportionally with $N$.
\end{itemize}
These properties mainly are extracted from a Hamiltonian
formulation for baryons as a bound-state of $N$ quarks. Based on
an approximation to approach the $N$-body problem (Hartree
approximation), the above properties can be justified for baryons
at large-$N$.

Here we try to work out the Hamiltonian formulation, and then the
above mentioned properties are followed by the same reasoning of
\cite{witN} \footnote{Because we have considered the D0-branes in
Light-Cone Frame, for $p^+=m_0\gg l_s^{-1}$, the heavy quark theory
of \cite{witN} is a good approximation for the transverse
dynamics of D0-branes.}.

In Sec.4 the effective theory for D0-branes were obtained to be
\bea
S=\int dt \bigg(\frac{1}{2} m_0\sum_{(a)=1}^{N} \dot{\vec{r}}_{(a)}^2-
4 \pi (\frac{d-1}{2}) \sum_{(a)>(b)=1}^{N}
\frac{|\vec{r}_{(a)}-\vec{r}_{(b)}|}{2\pi\alpha'}\bigg).
\eea
Also we have found the relation between the $\alpha'$ parameter
and the coupling constant of gauge theory by comparing it to LClgt, namely
$\frac{1}{\alpha'}\sim \frac{\gym^2}{a^2}$
where $a$ is the lattice spacing parameter. It is
known that it is more convenient to replace the coupling constant
by $\frac{\gym}{\sqrt{N}}$ at large-$N$ \cite{witN}.
So the action in terms of new parameters is
\bea
S=\int dt \bigg(\frac{1}{2} m_0\sum_{(a)=1}^{N} \dot{\vec{r}}_{(a)}^2-
4 \pi (\frac{d-1}{2}) \frac{\gym^2}{a^2}\frac{1}{N} \sum_{(a),(b)=1}^{N}
|\vec{r}_{(a)}-\vec{r}_{(b)}|\bigg),
\eea
and the associated Hamiltonian is the same used in
\cite{witN} except for the potential term, which is Coulomb one
there.

Here we just check the mass of baryons at large-$N$. The kinetic
term of c.m., $\frac{\vec{P}^2}{Nm_0}$,
grows with $N$, and the net potential for each D0-brane takes
a factor $\frac{1}{2}N(N-1)$ due to pair interactions. So the potential term
at large-$N$ grows like
\bea
\frac{1}{2}N(N-1)\frac{\gym^2}{N}\sim N.
\eea
It results that the energy grows as $E\sim N$ at large-$N$.
From the point of view of Light-Cone Frame the energy is $P^-$. The
total longitudinal momentum of this bound-state is $P^+=Np^+$, where
$p^+=m_0$ is the longitudinal momentum of one D0-brane. Consequently, the invariant
mass $M$ is
\bea
M^2=2P^+P^--\vec{P}^2\sim N^2\Rightarrow M\sim N.
\eea

\subsection{Quarks, Gauge Theory And Schwartzschild Solutions Of Gravity}
D$p$-branes are $p$ dimensional Schwartzschild
solutions of low energy effective
field theories of string theories
\footnote{In super string theories, they are charged solutions under
$p+1$-form field.}.
So any proposal for equivalence between them and quarks,
or at least between their dynamics and quarks dynamics, may need
justification at first. Here we recall some string theoretic
related issues shortly, and also try to present
(maybe) a  non-string theoretic related feature then.

As mentioned, D-branes are gravity solutions. On the other hand,
it is known that the dynamics of these objects are captured by a
gauge theory. It is one of the closest connections between
gauge theories and gravity, which has been revealed by string
theory. Through this relation between the dynamics of an extended
object and a gauge theory, many studies have been done to develop
understanding of gauge theory dynamics. One of the recent progresses
in this area is the adS/CFT correspondence \cite{adscft}, to
relate gauge theory dynamics at large 't Hooft coupling
($\lambda=\gym^2N$) to gravity in the anti-de Sitter background.

The relation between gauge theory and gravity is also studied at
the level of equations of motion. Both gravity and non-Abelian
gauge theories, though in different orders, have nonlinear
equations of motion. It is discovered that both pure gauge
theories and gauge theories with matter have Schwartzschild-like
solutions \cite{lunev}\cite{sing1}\cite{sing2}\cite{yoshida}. By
Schwartzschild-like we mean the similarity between ``connections''
in gauge theories (known as gauge fields $A_{(a)}^\mu$) and
gravity (known as connection coefficients
$\Gamma^\alpha_{\beta\gamma}$). In the case of $SU(2)$ gauge
theory with massless scalar matter field the solution is
found to be \cite{sing1}
\bea\label{bgs}
A^{(a)}_i&=&
\epsilon_{(a)ij}\frac{r^j}{\gym r^2}[1-K(r)],\nonumber\\
A^{(a)}_0&=& \frac{r^{(a)}}{\gym r^2}J(r),\nonumber\\
\phi^{(a)}&=& \frac{r^{(a)}}{\gym r^2}H(r),
\eea
with
\bea
K(r)&=&\frac{Cr}{1-Cr},\;\;\;\; J(r)=\frac{B}{1-Cr},\nonumber\\
H(r)&=&\frac{A}{1-Cr},\;\;\;\;\;{\rm with}\;\;A^2-B^2=1.
\eea
The gauge fields behaviour is comparable with connection coefficients
in Schwartzschild solution as
\bea
\Gamma^t_{rt}=\frac{K}{2r}\frac{1}{r-K},\;\;\;\;\;
\Gamma^r_{rr}=-\frac{K}{2r}\frac{1}{r-K},\;\;\;\;\; {\rm
with}\;\;K=2GM.
\eea
Here we just review some properties of the
solution (\ref{bgs}) \cite{sing1}. First, both the gauge and
scalar fields are singular at the radius $r_0=C^{-1}$. Further, by
calculating electric and magnetic fields one sees that both are
singular at $r_0$. Therefore a particle, which carries an $SU(2)$
charge, becomes confined if it crosses into the region
$r<r_0$. The singularity of field strengths at $r_0$ here is
different from that of gravity Schwartzschild solution, which can
be removed by a suitable choice of coordinates. Based on this
picture of confinement of a charge in $r<r_0$ region, a model for
confinement of gauge theories has been presented in
\cite{yoshida}.

Also the solution have monopole magnetic charge. This can be seen
from the generalized 't Hooft's field strength:
\bea
{\cal{F}_{\mu\nu}}=\partial_\mu(\hphi^{(a)}W^{(a)}_\nu)
-\partial_\nu(\hphi^{(a)}W^{(a)}_\mu)-
\frac{1}{\gym}\epsilon^{(a)(b)(c)}\hphi^{(a)}(\partial_\mu
\hphi^{(b)})(\partial_\nu\hphi^{(c)}),
\eea
with $\hat\phi^{(a)}\equiv \phi^{(a)} (\phi^{(b)}\phi^{(b)})^{-1/2}$.
Hence, for magnetic field we find
\bea
{\cal{B}}_i=\frac{1}{2}\epsilon_{ijk}{\cal{F}}_{ij}=-\frac{r^i}
{\gym r^3},
\eea
which is the magnetic field of a point monopole
with charge $-4\pi/\gym$. One can also find the
electric field:
\bea
{\cal{E}}_i=-{\cal{F}}_{0i}=\frac{r_i}{\gym
r}\frac{d}{dr}\frac{J(r)}{r} =\frac{B(2Cr-1)r_i}{\gym
r^3(1-Cr)^2},
\eea
which at $r\rightarrow \infty$ does not have
the behaviour of Prasad-Sommerfield's solution ($1/r^2$); and the
interpretation of a net electric charge near origin is impossible.
So this solution seems more like a magnetic monopole, and its
relation to a ``quark'' (a source of electric field) is
out of reach; but here the idea of Mantonen-Olive duality, which
changes the role of solitonic solutions with the fundamental
objects seems considerable.

\subsection{Why Non-Commutativity?}
One of the most interesting aspects of D-branes is the
non-commutativity of their relative coordinates.  If the model of
this paper has some relation with Nature, the question will be
about a possible justification for this non-commutativity. To
resolve this question one may consider the following prescription:
{\it The structure of space-time has to be in correspondence and
consistent with the propagation of fields}. In this way one finds
the space-time coordinates as $X_\mu$ 4-vector which behaves like
electromagnetic field $A_\mu$ 4-vector (spin 1) under the boost
transformations. This is just the same idea of special relativity
to change the concept of space-time to be consistent with the
Maxwell equations.

Also in this way supersymmetry is a natural continuation of the
special relativity program: Adding spin $\frac{1}{2}$ sector to the
coordinates of space-time, as the representative of the fermions of
nature.  This leads one to the space-time formulation of the supersymmetric
theories, and in the same way ferminos are introduced into the
bosonic string theory.

Now, what may be modified if nature has non-Abelian
(non-commutative) gauge fields? In the present nature non-Abelian
gauge fields can not make spatially long coherent states; they are
confined or too heavy.  But the picture may be changed inside a
hadron. In fact recent developments of string theories sound this
change and it is understood that non-commutative coordinates and
non-Abelian gauge fields are two sides of one coin.  As we discussed,
the interaction between D-branes is the result of
path-integrations over fluctuations of the non-commutative parts
of coordinates. It means that in this picture ``non-commutative''
fluctuations of space-time are the source of ``non-Abelian''
interactions. This picture may justify involving the
non-commutative coordinates in a study of bound-states of quarks
and gluons. One may summarize this idea as in the below table.

\vspace{.3cm}
\begin{center}
\begin{tabular}{|c|c|c|}
\hline Field & Space-Time Coordinate & Theory \\ \hline
\hline Photon $A^\mu$ & $X^\mu$ & Electrodynamics (and QED) \\
\hline Fermion  $\psi$ &  $\theta$, $\bar\theta$ & Supersymmetric\\
\hline Gluon $A^\mu_{(a)}$ &  $X^\mu_{(a)}$ & Chromodymamics (and QCD)\\
\hline
\end{tabular}
\end{center}
\vspace{.3cm}

{\bf Acknowledgement}

I am grateful to S. Parvizi for useful discussions and also comments on the
manuscript, and to M. Chaichian for his comment on large-$N$ consideration and
discussions on scattering amplitudes.  M.M. Sheikh-Jabbari's comments on the
revised version are deeply acknowledged. Finally I am grateful to the referee
of EPJ.C., for his/her crucial comments. 

\appendix
\section{Particle-Electrodynamics In Light-Cone Frame}
\setcounter{equation}{0}
We just follow the steps of \cite{lc2} in
going to Light-Cone Frame. The classical action is
\bea
S=-m\int_1^2 d\tau \sqrt{{\dot x}^2}+q\int A_\mu \dot{x}^\mu
d\tau, \eea with the momentum as \bea p^\mu\equiv -\frac{\partial
L}{\partial \dot{x}_\mu}=m
\frac{\dot{x}^\mu}{\sqrt{\dot{x}^2}}-qA^\mu.
\eea
Consequently one finds the constraints for momenta and canonical
Hamiltonian
\bea
(p^\mu+qA^\mu)(p_\mu+qA_\mu)=m^2,\\ H_c=-p_\mu\dot{x}^\mu-L\equiv
0.
\eea
The total Hamiltonian will be found to be
\bea
H_t=\lambda((p^\mu+qA^\mu)^2-m^2),
\eea
with $\lambda$ as Lagrange
multiplier, and canonical Poisson bracket as $\{x^\mu,
p^\nu\}=-\eta^{\mu\nu}$. So one finds that the dynamics has gauge
symmetry (reparametrization invariance) and to find the Lagrange
multiplier one should fix the gauge, by condition as
$\chi(x;\tau)\equiv 0$. Preserving gauge fixing during the time
gives
\bea
\dot\chi=0=\frac{\partial \chi}{\partial \tau}+ \{\chi, H_t\},
\eea
which gives
\bea
\lambda=-\{\chi,\theta\}^{-1}
\frac{\partial \chi}{\partial \tau},\;\;\;
\;\;\;\theta\equiv(p^\mu+qA^\mu)^2-m^2.
\eea
The Light-Cone gauge
fixing is $\chi=\tau-x^+=0$, and also by adding the gauge fixing
for the gauge field as $A^+=0$ \cite{lc1}, one finds for momentum
conjugate of time ($x^+$), i.e. Hamiltonian:
\bea
H=p^-=\frac{(\vec p+q\vec A)^2}{2p^+}-qA^-,
\eea
which here we
have assumed $m=0$. By taking $p^+$ as the Newtonian mass $m_0$ in
the transverse directions and $A^-$ as $A_0$, one gets the
Lorentz's equation of motion (\ref{lf}) by this Hamiltonian.


\end{document}